\documentclass[twocolumn,aps,prd,preprintnumbers]{revtex4}

\usepackage{graphicx}
\usepackage{dcolumn}
\usepackage{bm}
\usepackage{subfigure}
\usepackage{marvosym}
\usepackage{textcomp}
\usepackage{threeparttable}
\usepackage{booktabs}
\usepackage{setspace}
\usepackage{latexsym}
\usepackage{mciteplus}
\renewcommand{\TPTtagStyle}%
{\normalsize\textit}

\begin{document}

\title{Elliptic flow from thermal photons with magnetic field in holography}
\author{Berndt M\"uller$^{1,2}$\footnote{muller@phy.duke.edu}, Shang-Yu Wu$^{3,4,5}$\footnote{loganwu@gmail.com}, Di-Lun Yang$^1$\footnote{dy29@phy.duke.edu}}
\affiliation{
$^1$Department of Physics, Duke University, Durham, North Carolina 27708, USA.\\
$^2$Brookhaven National Laboratory, Upton, NY 11973, USA.\\
$^3$Institute of physics, National Chiao Tung University, Hsinchu 300, Taiwan.\\
$^4$National Center for Theoretical Science, Hsinchu, Taiwan.\\
$^5$Yau Shing Tung Center, National Chiao Tung University, Hsinchu, Taiwan}%
\date{\today}
\begin{abstract}
We compute the elliptic flow $v_2$ of thermal photons in a strongly coupled plasma with constant magnetic field via gauge/gravity duality. The D3/D7 embedding is applied to generate the contributions from massive quarks. By considering the cases in 2+1 flavor SYM analogous to the photon production in QGP, we obtain the thermal-photon $v_2$, which is qualitatively consistent with the direct-photon $v_2$ measured in RHIC at intermediate energy. However, due to the simplified setup, the thermal-photon $v_2$ in our model should be regarded as the upper bound for the $v_2$ generated by solely magnetic field in the strongly coupled scenario.   
\end{abstract}

\maketitle
The elliptic flow $v_2$ characterizes the momentum anisotropy of produced particles in heavy ion collisions.
The recent observations from RHIC and LHC revealed surprising results, where the large elliptic flow of direct photons has been measured\cite{Adare:2011zr,Lohner:2012ct}. Unlike the hadronic flow, the large flow of direct photons is unexpected since the high-energy photons are presumed to be generated in early times, where the initial flow should be relatively small compared to the flow built up by hydrodynamics. The anisotropy flow of thermal photons with viscous hydrodynamics has been recently reported in \cite{Dion:2011pp,Shen:2013cca}.
In theory, novel mechanisms should be introduced to break the azimuthal symmetry of photon production. The magnetic field led by colliding nuclei has been recently considered as one of possible candidates to bring about the large flow. In the weakly coupled scenario, the photon production with magnetic field has been studied in a variety of approaches\cite{Tuchin:2010gx,Tuchin:2012mf,Basar:2012bp,Fukushima:2012fg,Bzdak:2012fr}. 
Other mechanism such as the synchrotron radiation from the interaction of escaping quarks with the collective confining color field has been proposed in \cite{Goloviznin:2012dy}.

However, in the strongly coupled scenario, the perturbative approaches may not be applied. The AdS/CFT correspondence\cite{Maldacena:1997re,Witten:1998qj,Gubser:1998bc,Aharony:1999ti,Witten:1998zw}, a holographic duality between a strongly coupled $\mathcal{N}=4$ Super Yang-Mills(SYM) theory and a classical supergravity in the asymptotic $AdS_5\times S^5$ background in the limit of large $N_c$ and strong t'Hooft coupling, is thus introduced to handle nonperturbative problems. Although the precise dual of QCD is unknown, the SYM and QCD may share same qualitative features in the strongly coupled regime at finite temperature. The thermal photon production from adjoint matters in the holographic dual was initiated by \cite{CaronHuot:2006te} and then the one from fundamental matters was investigated in \cite{Mateos:2007yp}. The relevant studies of thermal photons have been generalized to the QCD duals\cite{Parnachev:2006ev,Jo:2010sg,PhysRevD.86.026003} and the SYM duals with the intermediate coupling\cite{Hassanain:2011ce} or with pressure anisotropy\cite{Rebhan:2011ke,Patino:2012py}. On the other hand, the computations of prompt photons and dileptons generated in early times via holography have been analyzed as well\cite{Baier:2012tc,Baier:2012ax,Steineder:2012si,Steineder:2013ana}.

Motivated by the anomalous flow of direct photons in heavy ion collisions, the thermal photon production with constant magnetic field in holography have been studied \cite{Mamo:2012zq,PhysRevD.87.026005,Yee:2013qma,Wu:2013qja,Arciniega:2013dqa}. In \cite{PhysRevD.87.026005}, it is shown that the photon production perpendicular to the magnetic field in D3/D7 and D4/D6 embeddings with massless quarks is enhanced. In \cite{Yee:2013qma}, the photon $v_2$ is computed in the framework of Sakai-Sugimoto model\cite{Sakai:2004cn}. The back reacted geometry in the presence of magnetic field may become anisotropic, which also results in an enhancement of photon production\cite{Arciniega:2013dqa}.
Furthermore, it is intriguing that the resonance in photon spectra from the meson-photon transition may lead to a mild peak of $v_2$ as pointed out in \cite{Wu:2013qja} when the photon production from massive quarks in D3/D7 embeddings is considered. To manifest the influence of the resonance on the elliptic flow, we will compute the $v_2$ of thermal photons in D3/D7 embeddings with constant magnetic field and incorporate the contributions from both massless and massive quarks.             
  
\begin{figure}[h]
\begin{center}
{\includegraphics[width=7.5cm,height=5cm,clip]{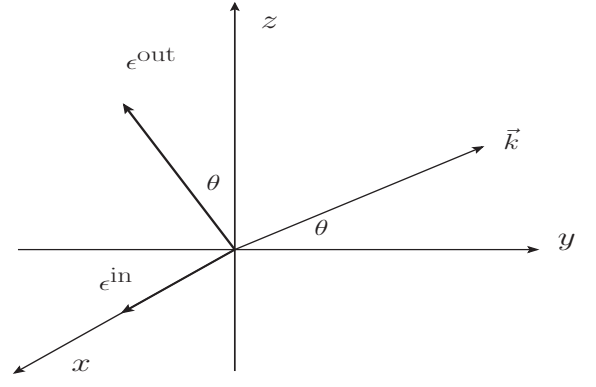}}
\caption{The coordinates of the system, where the magnetic field points along the z axis and the x axis is parallel to the beam direction. The $\vec{k}$ denotes the momentum of emitted photons and $\theta$ denotes the angle between the momentum and the x-y plane as the reaction plane; $\epsilon^{out}$ and $\epsilon^{in}$ represent the out-plane and in-plane polarizations, respectively.}\label{coord}
\end{center}
\end{figure}

Our setup is illustrated in Fig.\ref{coord}, where the magnetic field is along the $z$ direction and two types of polarizations $\epsilon^{\mbox{in}}$ and $\epsilon^{\mbox{out}}$ are considered. The four momentum of photons is written as $k=(-\omega,0,q_y,q_z)$, where $q_y=\omega\cos\theta$ and $q_z=\omega\sin\theta$. We will generalize the computations in the isotropic case of \cite{Wu:2013qja} to the photon production with arbitrary angle $\theta$. We will take the quenched approximation by assuming $N_f\ll N_c$, where $N_f$ denotes the number of flavors, and neglect the modification of flavor probe branes to the background geometry. The induced metric on D7 brane in the AdS-Schwarzschild background reads\cite{Karch:2002sh,Mateos:2006nu,Hoyos:2006gb}
\begin{eqnarray}\label{D7brane}
ds^2_{D7}&=&\frac{1}{u^2}\left(-f(u)dt^2+dx^2+dy^2+dz^2\right)\\ \nonumber
&+&\frac{1-\psi(u)^2+u^2f(u)\psi(u)'^2}{u^2f(u)(1-\psi(u)^2)}du^2+(1-\psi(u)^2)d\Omega_3^2,
\end{eqnarray}
where $\sqrt{1-\psi(u)^2}$ represents the radius of the internal $S^3$ wrapped by the D7 branes and $f(u)=1-u^4/u_h^4$ denotes the blackening function for $u_h$ being the even horizon. Here we set the AdS radius $L=(4\pi g_sN_cl_s^4)^{1/4}=1$. The temperature of the medium is determined by $\pi T=u_h^{-1}$. For convenience, we will further set $u_h=1$ in computations. We then turn on the worldvolume U(1) gauge field $2\pi l_s^2A_y=u_h^2B_z x$ coupled to the D7 branes, which generates constant magnetic field $eB=u_h^2B_{z}/(2\pi l_s^2)=B_{z}\pi^{-1}\sqrt{\lambda/2}$ along the $z$ direction, where $\lambda=g_{YM}^2N_c=2\pi g_sN_c$ denotes the t'Hooft coupling. To further introduce the electromagnetic currents, we should perturb the D7 branes with worldvolume gauge fields.  
The relevant part of the DBI action now takes the form,
\begin{eqnarray}\nonumber
S=-K_{D7}\int dtd^3\vec{x}duF^2\frac{(1-\psi^2)}{u^5}
(1+B_z^2u^4)^{1/2}\\
\times(1-\psi^2+u^2f\psi^{\prime 2})^{1/2},
\end{eqnarray}
where $F=dA$ is the worldvolume field strength from perturbation and   $T_{D7}=(2\pi l_s)^{-7} (g_sl_s)^{-1}$ is the D7-brane string tension for $K_{D7}=N_fT_{D7}(\pi l_s)^2\Omega_3$. In black hole embeddings corresponding to the deconfined phase, the field equation of $\psi$ in the DBI action with $F=0$ can be numerically solved by imposing the proper boundary conditions near the horizon \cite{Hoyos:2006gb}, $\psi(u_h)=\psi_0$ and $\psi'(u_h)=(-3u^{-2}\psi/f')|_{u=u_h}$. The asymptotic solution of $\psi(u)$ near the boundary behaves as
\begin{eqnarray}\label{psiboundary}
\psi(u)=m\frac{u}{2^{1/2}u_h}+c\frac{u^3}{2^{3/2}u_h^3}+\dots,
\end{eqnarray}
where the dimensionless coefficients $m$ and $c$ are related to the magnitudes of quark mass and condensate through \cite{Mateos:2006nu,Mateos:2007vn}
\begin{eqnarray}\label{mandc}
M_q&=&\frac{m}{2^{3/2}\pi l_s^2u_h}=\frac{\sqrt{\lambda}T}{2}m,\\\nonumber
\langle\mathcal{O}\rangle&=&-2^{3/2}\pi^3l_s^2N_fT_{D7}u_h^{-3}c=
-\frac{1}{8}\sqrt{\lambda}N_fN_cT^3c.
\end{eqnarray}

In the presence of gauge fields, the DBI action then gives rise to Maxwell equations 
\begin{eqnarray}
\partial_{\mu}(\sqrt{-\mbox{det}(G_{\mu\nu})}G^{\mu\alpha}G^{\nu\beta}F_{\alpha\beta})=0,
\end{eqnarray}
where the diagonal terms of the induced metric read
\begin{eqnarray}\nonumber
G^{tt}&=&-\frac{u^2}{f(u)},\quad G^{xx}=G^{yy}=\frac{u^2}{1+B_z^2u^4},\\
G^{zz}&=&u^2, \quad G^{uu}=\frac{u^2f(u)(1-\psi^2)}{1-\psi^2+u^2f(u)\psi'^2}.
\end{eqnarray}
To compute the spectral functions, it is more convenient to convert the field equations into gauge invariant forms. For the in-plane polarization $\epsilon_T=\epsilon^{in}=\epsilon_x$, the computation is straightforward. By taking $E_x=\omega A_x$ in momentum space, we have to solve only one field equation,
\begin{eqnarray}\label{invin}
E''_x+(\log(\sqrt{-G}G^{uu}G^{xx}))'E'_x-\frac{\bar{k}^2}{G^{uu}}E_x=0,
\end{eqnarray}
where $G=\mbox{det}(G_{\mu\nu})$ and $\bar{k}^2=G^{tt}w^2+G^{yy}q_y^2+G^{zz}q_z^2$. For the out-plane polarization $\epsilon_T=\epsilon^{out}$, we have to consider coupled equations.
By implementing the relation $q_yA_y+q_zA_z=0$ as shown in Fig.\ref{coord}, the field equations can be written into the gauge-invariant forms as
\begin{eqnarray}\label{invout}\nonumber
&E''_z&+\left[(\log(\sqrt{-G}G^{uu}G^{zz}))'+\frac{G^{zz}q_z^2}{\bar{k}^2}\left(\log\left(\frac{G^{tt}}{G^{zz}}\right)\right)'\right]E_z'\\\nonumber
&+&\frac{q_yq_zG^{yy}}{\bar{k}^2}\left(\log\left(\frac{G^{tt}}{G^{zz}}\right)\right)'E_y'-\frac{\bar{k}^2}{G^{uu}}E_z=0,\\\nonumber
&E''_y&+\left[(\log(\sqrt{-G}G^{uu}G^{yy}))'+\frac{G^{yy}q_y^2}{\bar{k}^2}\left(\log\left(\frac{G^{tt}}{G^{yy}}\right)\right)'\right]E_y'\\
&+&\frac{q_yq_zG^{zz}}{\bar{k}^2}\left(\log\left(\frac{G^{tt}}{G^{yy}}\right)\right)'E_z'-\frac{\bar{k}^2}{G^{uu}}E_y=0,
\end{eqnarray} 
where $E_{z(y)}=q_{z(y)}A_t+\omega A_{z(y)}$. The Maxwell equations in (\ref{invin}) and (\ref{invout}) can be solved numerically by imposing incoming-wave boundary conditions near the horizon\cite{CaronHuot:2006te}, 
where $\vec{E}(u)\sim (1-u^2/u_h^2)^{-\frac{i\omega}{4\pi T}}$.

Since $\bar{k}^2\approx-u^6\omega^2(1+B_z^2\cos^2\theta)$ near the boundary, eq.(\ref{invout}) reduces to
\begin{eqnarray}\label{EyEz}
(G^{yy}q_yE_y'+G^{zz}q_zE_z')_{u\rightarrow 0}=0.
\end{eqnarray}  
By utilizing the relation above, the near-boundary action can be simplified as
\begin{eqnarray}
\frac{-S_{\epsilon}}{2K_{D7}}
=\int\frac{d^4k}{(2\pi)^4}\frac{\sqrt{-G}G^{uu}}{\omega^2}G^{jj}E^*_jE_j',
\end{eqnarray}
where $j=x,y,z$.
We then evaluate the spectral density with the polarization $\epsilon_T$ via
\begin{eqnarray}\nonumber
\chi_{\epsilon_T}(k_0)&=&-4\mbox{Im}[\epsilon_T^{\mu}\epsilon_T^{\nu}C^R_{\mu\nu}(k)]
\\&=&-4\mbox{Im}\left[\lim_{u\rightarrow 0}\left(\omega^2\epsilon_{T\mu}\epsilon_{T\nu}\frac{\delta^2S_{\epsilon}}{\delta E^*_\mu E_{\nu}}\right)\right],
\end{eqnarray}
where $C^R_{\mu\nu}$ denotes the retarded correlator. For the in-plane polarization, we have
\begin{eqnarray}\nonumber
\frac{\chi_{\epsilon^{in}}}{8K_{D7}}&=&\frac{-1}{2K_{D7}}\mbox{Im}(C^{xx}_R)\\
&=&\mbox{Im}\left[\lim_{u\rightarrow 0}\left(\sqrt{-G}G^{uu}G^{xx}\frac{E_x'}{E_x}\right)\right].
\end{eqnarray}
For the out-plane polarization, we have
\begin{eqnarray}\label{chiout}
\frac{\chi_{\epsilon^{out}}}{8K_{D7}}&=&\frac{-1}{2K_{D7}}\mbox{Im}\Big[\sin^2\theta C^{yy}_R+\cos^2\theta C^{zz}_R\\\nonumber
&&-\cos\theta\sin\theta(C^{yz}_R+C^{zy}_R)\Big]\\\nonumber
&=&\mbox{Im}\left[\lim_{u\rightarrow 0}\left(\sqrt{-G}G^{uu}\left(G^{yy}\frac{E_y'}{E_y}+G^{zz}\frac{E_z'}{E_z}\right)\right)\right],
\end{eqnarray}
where we utilize (\ref{EyEz}) to derive the second equality above.
Solving the $E_z$ and $E_y$ for the out-plane polarization is more involved with the coupled equations, for which we discuss the technical details in the following. The procedure is similar to the computations in \cite{Patino:2012py}.  

Given that the out-plane solution is written in terms of the relevant bases as
\begin{eqnarray}
\vec{E}^{out}(u)=\vec{E}^{1}(u)+\vec{E}^{2}(u),
\end{eqnarray} 
where
$\vec{E}^{1}(u)=E^1_y\hat{y}(u)+E^1_z\hat{z}(u)$ and $\vec{E}^{2}(u)=E^2_y\hat{y}(u)+E^2_z\hat{z}(u)$, such bases should reduce to $\vec{E}^{1}(0)=E^1_y(0)\hat{y}$ and $\vec{E}^{2}(0)=E^2_z(0)\hat{z}$ on the boundary, which correspond to $E_y$ and $E_z$ in (\ref{chiout}). Since $A_t(0)=0$ on the boundary, the bases follow the constraint $E^1_y(0)/E^2_z(0)=-\tan\theta$. The task will be to find these relevant bases.

Presuming that $\vec{E}^a(u)=E^a_y(u)\hat{y}+E^a_z(u)\hat{z}$ and $\vec{E}^b(u)=E^b_y(u)\hat{y}+E^b_z(u)\hat{z}$ are two sets of incoming-wave solutions, the relevant bases should be formed by linear combinations of the them. We thus define 
\begin{eqnarray}\nonumber
\vec{E}^1(u)&=&a_1\vec{E}^a(u)+b_1\vec{E}^b(u),\\ \vec{E}^2(u)&=&a_2\vec{E}^a(u)+b_2\vec{E}^b(u).
\end{eqnarray}
The bases on the boundary then read
\begin{eqnarray}\nonumber
\vec{E}^1(0)&=&a_1\vec{E}^a(0)+b_1\vec{E}^b(0)=-E_0\sin\theta\hat{y},\\
\vec{E}^2(0)&=&a_2\vec{E}^a(0)+b_2\vec{E}^b(0)=E_0\cos\theta\hat{z},
\end{eqnarray}
where $E_0=|\vec{E}^{out}(0)|$. By solving the coupled equations above, we find
\begin{eqnarray}\nonumber\label{ab}
(a_1,b_1)&=&\frac{\left(-E^b_z(0)\sin\theta,E^a_z(0)\sin\theta\right)}{E^a_y(0)E^b_z(0)-E^b_y(0)E^a_z(0)},\\
(a_2,b_2)&=&\frac{\left(-E^b_y(0)\cos\theta,E^a_y(0)\cos\theta\right)}{E^a_y(0)E^b_z(0)-E^b_y(0)E^a_z(0)},
\end{eqnarray}
where we set $E_0=1$ since the retarded correlators are invariant for an arbitrary $E_0$. In practice, we could solve for two arbitrary incoming waves $\vec{E}^{a(b)}(u)$. Then by employing the coefficients shown in (\ref{ab}) to recombine these two solutions, we are able to derive $E_y(u)$ and $E_z(u)$ for the out-plane polarization.  

Finally, we may compute the elliptic flow $v_2$ for photon production. 
In the lab frame of heavy ion collisions, the four-momenta of photons can be parametrized as 
\begin{eqnarray}
k_{\mu}=(-k_T\cosh\tilde{y},k_T\sinh\tilde{y},k_T\cos\theta,k_T\sin\theta),
\end{eqnarray}
where $\tilde{y}$ denote the rapidity and $k_T$ denote the transverse momentum perpendicular to the beam direction $\hat{x}$. Notice that $\omega=k_T\cosh\tilde{y}\approx k_T$ at central rapidity$(\tilde{y}\approx 0)$, which reduces to our setup illustrated in Fig.\ref{coord}.
The elliptic flow $v_2$ is defined as
\begin{eqnarray}
v_2^{\gamma}(k_T,\tilde{y})=\frac{\int^{2\pi}_0d\theta\cos(2\theta)\frac{dN_{\gamma}}{d^2k_Td\tilde{y}}}{\int^{2\pi}_0d\theta\frac{dN_{\gamma}}{d^2k_Td\tilde{y}}},
\end{eqnarray}
where $N_{\gamma}$ is the total yield of the emitted photons.
In thermal equilibrium,
the differential emission rate per unit volume is given by
\begin{eqnarray}
\omega\frac{d\Gamma_{\gamma}(\epsilon_T)}{d^3k}=\frac{d\Gamma_{\gamma}(\epsilon_T)}{d^2k_Td\tilde{y}}
=\frac{1}{16\pi^3}\frac{\chi_{\epsilon_T}(k_0)}{(e^{\beta \omega}-1)}.
\end{eqnarray}
In general, we have to take four dimensional spacetime integral of the emission rate to obtain the yield of photons. In our setup, where the medium is static, the spacetime integral leads to a constant volume, which is irrelevant for $v_2$ here. The elliptic flow at central rapidity hence becomes
\begin{eqnarray}
v_2^{\gamma}(\omega,0)&=&\frac{\int^{2\pi}_0d\theta\cos(2\theta)\chi_{\epsilon_T}(k_0)}{\int^{2\pi}_0d\theta\chi_{\epsilon_T}(k_0)}.
\end{eqnarray}

All physical observables now will be scaled by temperature of the medium. We set $B_z=1(\pi T)^2$, which corresponds to $eB= 0.39$ GeV$^2$ in the regular scheme for $\lambda=6\pi$ and the average temperature of the SYM plasma $T=T_{\text{QGP}}= 200$ MeV. In an alternative scheme\cite{Gubser:2006qh}, $eB= 0.12$ GeV$^2$ for $\lambda=5.5$ and $T=3^{-1/4}T_{\text{QGP}}\approx 150$ MeV, where the temperature of SYM plasma is lower than that of QGP at fixed energy density.
In heavy ion collisions, the approximated magnitude of magnetic field is about the hadronic scale, $eB\approx m_{\pi}^2\approx 0.02$ GeV$^2$\cite{Tuchin:2013ie}. It turns out that the magnitude of magnetic field in the alternative scheme is close to the approximated value at RHIC. Even in the regular scheme, the magnitude of magnetic field in our model is not far from the approximated value. Hereafter we will make comparisons to QGP in the alternative scheme.

\begin{figure}[h]
\begin{center}
{\includegraphics[width=7.5cm,height=5cm,clip]{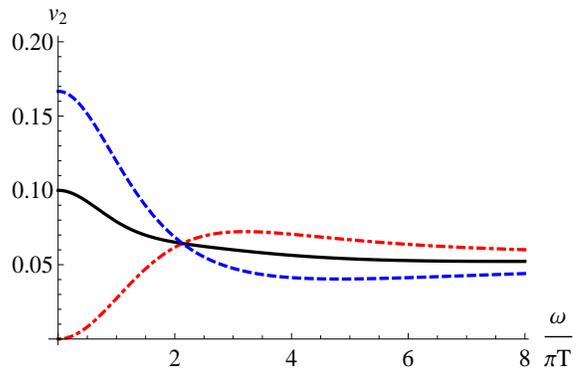}}
\caption{The red(dot-dashed) and blue(dashed) curves correspond to the $v_2$ of the photons with in-plane and out-plane polarizations, respectively. The black(solid) curve correspond to the one from the averaged emission rate of two types of polarizations. Here we consider the contribution from massless quarks at $B_z=1(\pi T)^2$.}\label{v2massless}
\end{center}
\end{figure}

\begin{figure}[h]
\begin{center}
{\includegraphics[width=7.5cm,height=5cm,clip]{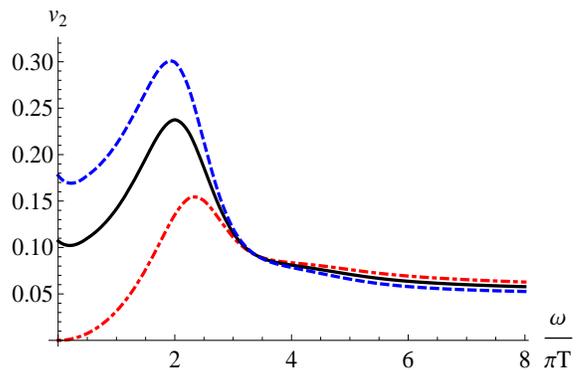}}
\caption{The colors correspond to the same cases as in Fig.\ref{v2massless}. Here we consider the contributions from solely the massive quarks with $m=1.143$ at $B_z=1(\pi T)^2$.}
\label{v2massive}
\end{center}
\end{figure}   

\begin{figure}[h]
\begin{center}
{\includegraphics[width=7.5cm,height=5cm,clip]{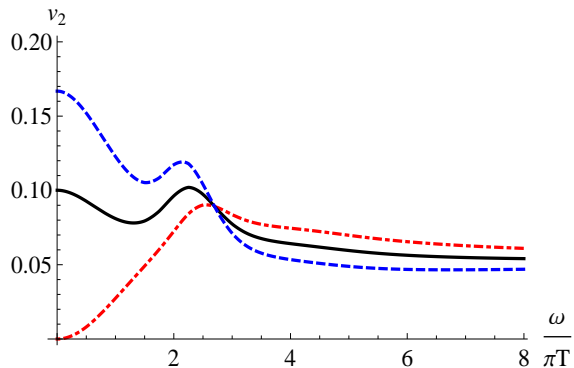}}
\caption{The colors correspond to the same cases as in Fig.\ref{v2massless}. Here we consider the contributions from both massless quarks and massive quarks with $m=1.143$ at $B_z=1(\pi T)^2$.}
\label{v2full}
\end{center}
\end{figure}   

We firstly consider the elliptic flow contributed from massless quarks, which corresponds to the trivial embedding$(\psi'=0)$. As shown in Fig.\ref{v2massless}, the presence of magnetic field results in nonzero $v_2$, while the $v_2$ remain featureless(without resonances). 
Here the averaged $v_2$ is obtained from the averaged emission rate of both the in-plane and out-plane polarizations.
Whereas quarks may receive mass correction at finite temperature, we should consider the contributions from massive quarks as well. In addition, at intermediate energy, the photon spectra from the massive quarks may lead to resonances originated from the decays of heavy mesons to lightlike photons\cite{Mateos:2007yp,CasalderreySolana:2008ne}, which bring about considerable contribution to the spectra. As indicated in \cite{Wu:2013qja}, the resonances in the presence of magnetic field depend on the moving directions of produced photons, which may generate prominent peaks in $v_2$.  
To incorporate the massive quarks, we choose $\psi_0\approx 0.95$, which is close to the critical embedding$(\psi_0\rightarrow 1)$. In fact, by further tuning $\psi_0$ up to one, the black hole embeddings may become unstable and multiple resonances will emerge in photon spectra similar to the scenarios in the absence of magnetic field\cite{Mateos:2007yp}. From (\ref{psiboundary}), we find $m=1.143$ for the solution of the massive quarks, which corresponds to the bare quark mass $M_q= 204$ MeV at the average RHIC temperature $T_{\text{QGP}}=200$ MeV in the alternative scheme. Due to the presence of magnetic field and the choice of the alternative scheme, the bare quark mass for the massive quark here is smaller than that in \cite{Mateos:2007yp,CasalderreySolana:2008ne} to generate the resonance. 
As shown in Fig.\ref{v2massive}, a mild peak emerges at intermediate energy for the photon $v_2$ contributed from solely the massive quarks.

In analogy to the thermal photon production in QGP, we may consider scenario in the 2+1 flavor SYM plasma. We sum over the photon emission rates from two massless quarks and that from the massive quark with $M_q= 204$ MeV to compute the $v_2$.
The results are shown in Fig.\ref{v2full}, where the resonances of $v_2$ are milder.   
In QGP, the regime in which the thermal photons make substantial contributions is around $p_T\approx 1\sim 4 $ GeV at central rapidity, where $p_T\approx\omega$ denotes the transverse momentum of direct photons.   
By rescaling $p_T$ with $\pi T_{\text{QGP}}$, such a regime corresponds to $\omega/(\pi T)\approx 1.5\sim 6$ in Fig.\ref{v2full} at $T_{\text{QGP}}=200$ MeV. It turns out that the $v_2$ in our holographic model resemble the RHIC data for the flow of direct photons at intermediate $p_T$\cite{Adare:2011zr}. 
Although the mass of the massive quark in our setup does not match that of the strange quark, the mass we introduce is not far from the scale of strange mesons. The resonances in our setup may suggest the transitions of strange mesons to photons in QGP in the presence of magnetic field. 
On the other hand, the resonance of $v_2$ coming from meson-photon transitions may not be subject to the strongly coupled scenario. In the weakly coupled approach such as \cite{Basar:2012bp}, where the finite-temperature corrections to the intermediate meson in the effective coupling is not considered, the photon production perpendicular to the magnetic field can be possibly enhanced provided that the thermal dispersion relation of the intermediate meson becomes lightlike.

\begin{figure}[h]
\begin{center}
{\includegraphics[width=7.5cm,height=5cm,clip]{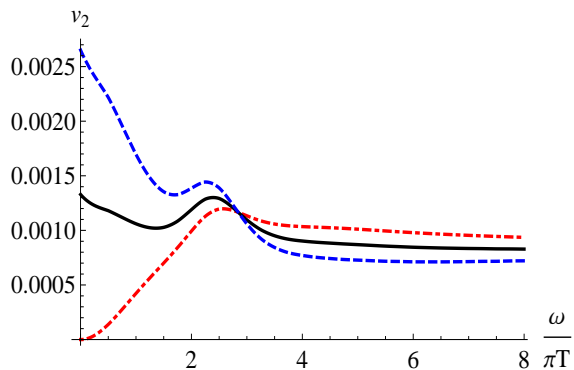}}
\caption{The colors correspond to the same cases as in Fig.\ref{v2massless}. Here we consider the contributions from both massless quarks and massive quarks with $m=1.307$ at $B_z=0.1(\pi T)^2$.}
\label{v2smallB}
\end{center}
\end{figure}

Finally, we mention the caveats when making comparisons between our holographic model and heavy ion collisions in reality except for the intrinsic difference between SYM theory and QCD. Firstly, the QGP undergoes time-dependent expansion, while the medium in our model is static in thermal equilibrium. Second, the magnetic field produced by colliding nuclei is time-dependent, which decay rapidly in early times. Although the influence of thermal quarks on the lifetime of magnetic field is controversial\cite{Tuchin:2013ie,McLerran:2013hla,Tuchin:2013apa}, the constant magnetic field in our model could overestimate the flow. 
According to \cite{Tuchin:2013ie}, the magnetic field decreases by a factor of $100$ between
the initial (0.1 fm/c)and final (5 fm/c) times in the presence of nonzero conductivity. As a simple approximation, we may assume that the magnetic field is described by a power-law drop-off, which results in $B(t)\sim 1/t^{1.2}$. By taking the initial and freeze-out temperature as $T_i=430$ MeV and $T_f=150$ MeV, we find the freeze-out time $\tau_f\sim 7$ fm as we set the thermalization time $\tau_{th}=0.3$ fm and average temperature $T_{avg}\sim 200$ MeV with the Bjorken hydrodynamics $T/T_i=(\tau_{th}/\tau)^{1/3}$. We than obtain the average magnetic field $B_{avg}\sim 0.1B_0$ with the setup above, where $B_0$ is the initial magnetic field. By utilizing the average magnetic field with the same t'Hooft coupling and average temperature, we find that the $v_2$ drop about $100$ times as shown in Fig.\ref{v2smallB}. 

\begin{figure}[h]
\begin{center}
{\includegraphics[width=7.5cm,height=5cm,clip]{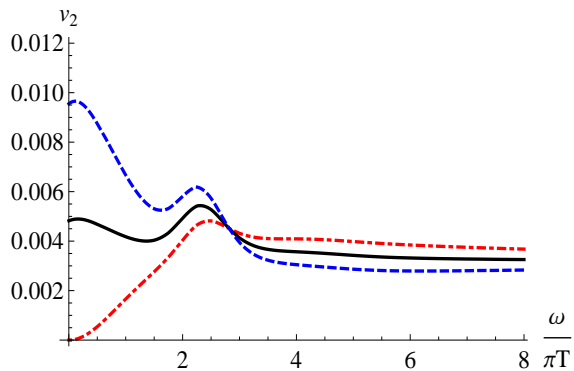}}
\caption{The colors correspond to the same cases as in Fig.\ref{v2massless}. Here we consider the contributions from both massless quarks and massive quarks with $m=1.3$ at $B_z=0.2(\pi T)^2$.}
\label{v2smallB2}
\end{center}
\end{figure}

Although the photon $v_2$ here can only be evaluated numerically, it is approximately proportional to $B_z^2$ for small $B_z$. As a result, we may as well consider the result with average $(eB)^2$. With the above approximation, we find $B^2_{avg}\sim 0.031B^2_0$ corresponding to $B_{avg}\sim 0.18B_0$. As shown in Fig.\ref{v2smallB2}, the $v_2$ with $B^2_{avg}\sim 0.031B^2_0$ drop $25$ times. 
However, as the nonlinear effect with large $B_z$ becomes more pronounced, the computation with average magnetic field may underestimate the contribution from such strong magnetic field in early times. It is thus desirable to incorporate time-dependent magnetic field in the setup as future work. On the other hand, it is also worthwhile to notice that the $v_2$ in our model is enhanced as we turn down the coupling with fixed magnetic field and temperature through the relation $eB=B_z\pi^{-1}\sqrt{\lambda/2}$.    

Since we choose the maximum magnetic field from its initial value, the $v_2$ obtained in our model should be regarded as the upper bound generated by solely magnetic field in the strongly coupled scenario. In reality, such a mechanism only yields partial contribution of the measured $v_2$. As shown in \cite{Shen:2013cca}, the viscous hydrodynamics also results in a substantial contribution to thermal-photon $v_2$. To construct full $v_2$ for thermal photons, both contributions from magnetic field and from viscous hydrodynamics should be taken into account.  
Furthermore, in the alternative scheme, the intermediate t'Hooft coupling is taken, where the  corrections from finite t'Hooft coupling in the gravity dual have to be considered. More explicitly, the next leading order correction is of $\mathcal{O}(\lambda^{-3/2})$. It is found in \cite{Hassanain:2011ce} that the photoemission rate increases as the coupling decreases in the absence of magnetic field when the $\mathcal{O}(\lambda^{-3/2})$ correction is included.

The authors thank S. Cao and G. Qin for useful discussions. This material is based upon work supported
by DOE grants DE-FG02-05ER41367 (B. M\"uller and D.~L.~Yang), the National Science Council (NSC
101-2811-M-009-015) and the Nation Center for Theoretical Science(102-2112-M-033-003-MY4), Taiwan (S.~Y.~Wu).  


\end{document}